\documentclass[twocolumn]{aa}
\usepackage{graphicx}
\usepackage{natbib}
\def\vev#1{\langle#1\rangle}

\def\pmb#1{\setbox0=\hbox{#1}%
 \kern-.025em\copy0\kern-\wd0
 \kern.05em\copy0\kern-\wd0l
 \kern-.025em\raise.0433em\box0}
\def\eg{{\it e.g.\ }}

\def\ie{{\it i.e.\ }}

\def\del{\delta}                
\def\delg{\delta_g}           
\def\sig{\sigma}              
\def\bl{b_L}                      
\def\hpc{$h^{-1}$Mpc }
\def\hpcp{$ h^{-1}$Mpc. }
\def\hpcv{$ h^{-1}$Mpc, }

\def\delt{\delta_{T}}
\def\tdelt{{\tilde\delta}_{T}}

\def\tdelf{{\tilde\delta}_{F}}

\def\tdelg{{\tilde \delta}_g}
\def\teps{{\tilde\epsilon}}

\def\z{$\,${\it z}$\,$}

\def\blr{{\bf r}}
\def\blk{{\bf k}}

\def\eg{{e.g.,~}}
\def\ie{{i.e.,~}}
\def\lsim{\raise0.3ex\hbox{$<$}\kern-0.75em{\lower0.65ex\hbox{$\sim$}}} 
\def\gsim{\raise0.3ex\hbox{$>$}\kern-0.75em{\lower0.65ex\hbox{$\sim$}}} 
\def\lesssim{\mathrel{\hbox{\rlap{\hbox{\lower4pt\hbox{$\sim$}}}\hbox{$<$}}}}
\def\gtrsim{\mathrel{\hbox{\rlap{\hbox{\lower4pt\hbox{$\sim$}}}\hbox{$>$}}}}
\def\sc1{\raise2.1ex\hbox{\tiny $r\!\!=\!\!4$}\kern-0.95em{\hbox{$=$}}}

\newcommand{\unit}[1]{\ifmmode \:\mbox{\rm #1}\else \mbox{#1}\fi}

\def\ltsima{$\; \buildrel < \over \sim \;$}
\def\simlt{\lower.5ex\hbox{\ltsima}}
\def\gtsima{$\; \buildrel > \over \sim \;$}
\def\simgt{\lower.5ex\hbox{\gtsima}}          

\def\sc{{\rm Science\ }}

\begin{document}

\title{
The VIMOS VLT Deep Survey}

\subtitle{Testing the gravitational instability paradigm at $z\sim 1$}

\author{
C. Marinoni \inst{1},
L. Guzzo \inst{2,3},
A. Cappi \inst{4},
O. Le F\`evre \inst{5},
A. Mazure \inst{5},
B. Meneux \inst{3},
A. Pollo \inst{6},
A. Iovino \inst{2},
H.J. McCracken \inst{7},
R. Scaramella \inst{8},
S. de la Torre \inst{5},
J. M. Virey \inst{1},
D. Bottini \inst{9},
B. Garilli \inst{9},
V. Le Brun \inst{5},
D. Maccagni \inst{9},
J.P. Picat \inst{10},
M. Scodeggio \inst{9},
L. Tresse  \inst{5},
G. Vettolani \inst{11},
A. Zanichelli \inst{11},
C. Adami \inst{5},
S. Arnouts \inst{5},
S. Bardelli \inst{4},
M. Bolzonella \inst{4},
S. Charlot \inst{7},
P. Ciliegi \inst{11},
T. Contini \inst{10},
S. Foucaud \inst{12},
P. Franzetti \inst{9},
I. Gavignaud \inst{13},
O. Ilbert \inst{14},
F. Lamareille \inst{10},
B. Marano \inst{15},
G. Mathez \inst{10},
R. Merighi \inst{4},
S. Paltani \inst{16,17},
R. Pell\`o \inst{10},
L. Pozzetti \inst{4},
M. Radovich \inst{18},
D. Vergani \inst{9},
G. Zamorani  \inst{4},
E. Zucca  \inst{4},
U. Abbas \inst{5},
M. Bondi \inst{11},
A. Bongiorno \inst{11},
J. Brinchmann \inst{19},
A. Buzzi \inst{1},
O. Cucciati \inst{5,20},
L. de Ravel \inst{5},
L. Gregorini \inst{11},
Y. Mellier \inst{7},
P. Merluzzi \inst{18},
E. Perez-Montero \inst{10},
P. Taxil \inst{1},
S. Temporin \inst{5},
C.J. Walcher \inst{7}
}


\institute{
Centre de Physique Th\'eorique, UMR 6207 CNRS-Universit\'e de Provence, Case 907, F-13288 Marseille, France
\and
INAF - Osservatorio Astronomico di Brera, via Brera 28, I-20121 Milano, Italy
\and
Max Planck Institut f\"ur extraterrestrische Physik, D-85741 Garching, Germany
\and
INAF - Osservatorio Astronomico di Bologna, via Ranzani 1, I-40127 Bologna, Italy
\and
Laboratoire d'Astrophysique de Marseille, UMR 6110, BP8 Traverse du Siphon, F-13012 Marseille, France
\and
Astronomical Observatory of the Jagiellonian University, ul Orla 171,
30-244 Krakow, Poland 
\and
Institut d'Astrophysique de Paris, UMR 7095, 98 bis Bvd. Arago,
F-75014 Paris, France 
\and
INAF - Osservatorio Astronomico di Roma, via Osservatorio 2, I-00040 Monteporzio Catone (Roma), Italy
\and
INAF - IASF,  Via Bassini 15, I-20133 Milano, Italy
\and
LAT - Observatoire Midi-Pyr\'en\'ees, UMR5572, 14 av. E. Belin, F-31400 Toulouse, France
\and
INAF - Istituto di Radio-Astronomia, Via Gobetti 101, I-40129 Bologna, Italy
\and
School of Physics \& Astronomy, University of Nottingham, University Park, Nottingham, NG72RD, UK
\and
Astrophysical Institute Potsdam, An der Sternwarte 16, D-14482, Potsdam, Germany 
\and
Institute for Astronomy, 2680 Woodlawn Dr., University of Hawaii, Honolulu, Hawaii, 96822, USA 
\and
Universit\`a di Bologna, Dipartimento di Astronomia, via Ranzani 1, I-40127 Bologna, Italy
\and
Integral Science Data Centre, ch. d'\'Ecogia 16, CH-1290, Versoix, Switzerland 
\and
Geneva Observatory, ch. des Maillettes 51, CH-1290, Sauverny, Switzerland 
\and
INAF - Osservatorio Astronomico di Capodimonte, via Moiariello 16, I-80131 Napoli, Italy
\and
Centro de Astrof{\'{i}}sica da Universidade do Porto, Rua das Estrelas, P-4150-762, Porto, Portugal 
\and
Universit\`a di Milano-Bicocca, Dipartimento di Fisica, Piazza della scienza 3, I-20126 Milano, Italy
}

\authorrunning{Marinoni et al.}
\titlerunning{Testing Gravity at z$\sim 1$}


\abstract{

  We have reconstructed the three-dimensional density fluctuation maps
  to $z\sim 1.5$ using the distribution of galaxies observed in the
  VVDS-Deep survey.  We use this overdensity field to measure the
  evolution of the probability distribution function and its
  lower-order moments over the redshift interval 0.7$<$\z$<$1.5. We
  apply a self-consistent reconstruction scheme which includes a
  complete non-linear description of galaxy biasing and which has been
  throughly tested on realistic mock samples. We find that the
  variance and skewness of the galaxy distribution evolve over this
  redshift interval in a way that is remarkably consistent with
  predictions of first- and second-order perturbation theory.  This
  finding confirms the standard gravitational instability paradigm
  over nearly 9 Gyrs of cosmic time and demonstrates the importance of
  accounting for the non-linear component of galaxy biasing to
  consistently reproduce the higher-order moments of the galaxy
  distribution and their evolution.

}
\maketitle

\section{Introduction}

According to Thomas Wright mapping the Cosmos on the very largest
scales is about gaining {\it ``a partial View of Immensity, or without
much Impropriety perhaps, a finite View of Infinity''}\footnote{ {\it
An Original Theory of the Universe} (1750, 9th letter, Plate XXXI).}.
Unfortunately, charting the cosmic territory beyond our local volume
into the distant Universe is observationally challenging. Until
recently, our understanding of the large-scale organisation of
galaxies at $z \gtrsim 0.2$ had to rely on the predictions of
numerical simulations in the framework of the rather successful cold
dark matter model \citep[\eg][]{spri05}.

Within this scenario, which has now developed into the leading
theoretical paradigm for the formation of structures in the Universe,
structures grow from weak, dark-matter density fluctuations present in
the otherwise homogeneous and rapidly expanding early universe.  The
standard version of the model incorporates the assumption that this
primordial, Gaussian-distributed fluctuations are amplified by
gravity, eventually turning into the rich structures we see today.

This picture in which gravity, as described by general relativity, is
the engine driving cosmic growth is generally referred to as the
gravitational instability paradigm (GIP). However plausible it may
seem, it is important to test its validity. In the local universe the
GIP paradigm has been shown to make sense of a vast amount of
independent observations on different spatial scales from galaxies to
superclusters of galaxies \citep[\eg][]{pea01,teg06}.  Deep redshift
surveys now allow us to test whether the predictions of this
assumption are also valid at earlier epochs.

In this paper we test the role of gravity in shaping density
inhomogeneities by using three-dimensional maps of the distribution of
visible matter revealed by the VIMOS-VLT Deep Survey over the large
redshift baseline $0<z<1.5$ (see Massey et al. 2007 for three
dimensional cartography of mass overdensities in the COSMOS field).  We
present first a qualitative picture of the large-scale organization of
remote cosmic structures, and then quantify the observed clustering by
computing the probability distribution functions (PDF) of galaxy
overdensities $\delta_g$. In this way, we trace how the amplitude and
spatial arrangement of galaxy fluctuations changes with cosmic time.
We explore the mechanisms governing this growth by comparing the time
evolution of the low-order moments of the galaxy PDF, ({\em i.e.}  the
{\it variance} amplitude $<\delta_g^2>$ and the {\it normalised
  skewness} $S_3=<\delta_g^3>_c/<\delta_g^2>^{2}$) with the
corresponding quantity theoretically predicted for matter fluctuations
in the linear and semi-linear perturbative regime. (Note that in the
following we shall often speak equivalently of the variance or of its
square root, i.e.  the {\it root mean square} amplitude
$<\delta_g^2>^{1/2}$ when referring to the second-order moment). This
provides a test of GIP-specific predictions at as-yet unexplored epochs
that are intermediate between the present era and the time of
decoupling.  Knowledge of the precise growth history of density
inhomogeneities provides also a way to test the theory of
gravitation~\citep[\eg][]{lin05}.

In addition to the statistical approach presented in this paper, we
have recently addressed this same issue also from a dynamical point of
view. We have used linear redshift-space distortions in the VVDS-{\it
Wide} data to measure the growth rate of matter fluctuations at
$z\sim 0.8$ \citep{nature07}.  This approach offers promising prospects
for determining the cause of cosmic acceleration in the near future
\citep{lin07}.  The work presented here is also complemented by a
parallel paper (Cappi et al. 2008) in which we study the behavior of
the N-point correlation functions for this same sample. Higher-order
galaxy correlation functions are known to display a hierarchical
scaling as a function of the variance of the count distribution (\eg
Peebles 1980). In the same spirit, we use this scaling to test the
standard assumption of evolution under gravitational instability of an
initially Gaussian distribution of density fluctuations.

The paper is organised as follows: in \S 2 we briefly describe the
first-epoch VVDS data sample. In \S 3 we present 3D overdensity maps
from the galaxy distribution in the VVDS to $z \sim 1.5$; we then
characterise the evolution of galaxy fluctuations with cosmic epoch by
computing their PDF in two redshift slices.  In \S 4 we compare the
observed redshift evolution of the low-order moments (i.e. variance and
skewness) of the PDF of the galaxy fluctuations with linear and
semi-linear theoretical predictions of the Gravitational Instability
Paradigm.  Conclusions are presented in \S 5.

The coherent cosmological picture emerging from independent
observations and analysis motivates us to present our results in the
context of a $\Lambda$CDM cosmological model with $\Omega_m=0.3$ and
$\Omega_{\Lambda}=0.7$.  Throughout, the Hubble constant is
parameterised via $h=H_{0}/100$.  All magnitudes in this paper are in
the AB system (Oke \& Gunn 1983), and from now on we will drop the
suffix AB.

\section{The First-Epoch VVDS-Deep  Redshift Sample}
\label{data}

The primary observational goal of the VIMOS-VLT Redshift Survey as well
as the survey strategy and first-epoch observations in the VVDS-0226-04
field (from now on simply VVDS-02h) are presented by \citet{lef05}.
Here it is enough to stress that, in order to minimise selection
biases, the VVDS-Deep survey has been conceived as a purely
flux-limited ($17.5\leq I \leq24$) survey, \ie no target pre-selection
according to colors or compactness is used.  Stars and QSOs have been
{\it a-posteriori} removed from the final redshift sample.  Photometric
data in this field are complete and free from surface brightness
selection effects, down to the limiting magnitude $I_{AB}$=24
\citep{mcc03}.  Spectroscopic observations were carried out using the
VIMOS multi-object spectrograph using one arcsecond wide slits and the
LRRed grism which covers the spectral range $5500<\lambda(\AA)<9400$
with an effective spectral resolution $R\sim 227$ at $\lambda=7500\AA$.
The {\it rms} accuracy in the redshift measurements is $\sim$275 km/s.
Details on the observations and data reduction are given in
\citet{lef04} and in \citet{lef05}.

The VVDS-02h data sample extends over an area of 0.7$\times$0.7 sq.deg
(which was targeted according to a one, two or four passes strategy, \ie
giving to any single galaxy in the field one, two or four chances to be
targeted by VIMOS masks (see fig. 12 in \citet{lef05}) has a median
depth of about \z$\sim$0.76.  It contains 6582 galaxies with secure
redshifts (\ie redshift determined with a quality flag$\ge$2 (see
\citet{lef05})) and probes a comoving volume (up to $z=1.5$) of nearly
$1.5\cdot 10^6 h^{-3}$ Mpc$^{3}$.  This volume has transverse
dimensions $\sim 37 \times 37$ \hpc at $z=1.5$ and extends over a
comoving length of 3060 \hpc in the radial direction.

For the statistical analysis presented in this paper, we first define a
sub-sample (VVDS-02h-4) including galaxies with redshift \z$<$1.5 and
over the sky region (0.4$\times$0.4 deg$^2$) that was repeatedly
covered by four independent VIMOS observations in each point.  Even if
measured redshifts in the VVDS reach up to \z$\sim$5 and cover a wider
area, these conservative limits bracket the range where we can sample
in a denser way the underlying galaxy distribution and, thus, minimise
biases in the reconstruction of the density field (see the analysis in
\S 4.1).  The VVDS-02h-4 subsample contains 3448 galaxies with secure
redshift (3001 with $0.4<z<1.5$), probes one-third of the total
VVDS-02h volume and it is characterised by a redshift sampling rate of
$\sim 30\%$ (i.e. on average about one over three galaxies with
magnitude $I_{AB}\leq$24 has a measured redshift).  This high spatial
sampling rate is a critical factor for minimising biases in the
reconstruction of the 3D density field of galaxies.  To optimise the
analysis of the associated probability density function, we further 
select only galaxies with absolute blue magnitude $M_B<-20+\log h$.  With this
selection, we define two nearly volume-limited sub-samples in the
redshift ranges $0.7<z<1.1$ and $1.1<z<1.5$ respectively. A discussion
of possible effects of galaxy evolution on our results is presented in
\S~4.3.

\section{The galaxy density field at high redshift}

The first large redshift surveys of the local Universe
\citep[e.g.][]{dav81,gh89,gh91,s92a,dac94} showed that galaxies have a
highly non-random spatial distribution and cluster in a hierarchical
fashion.  The corresponding three-dimensional maps reveal a complex
web-like network of thin, filamentary structures connecting centrally
condensed clusters of galaxies, punctuated by large, quasi-spherical,
low-density voids.  These structures are the outcome of more than 13
billion years of evolution of small-amplitude fluctuations that we see
reflected in the temperature anisotropy of the Cosmic Microwave
Background (CMB) at $z\simeq 1100$ \citep{sper07}.  Recent analyses
(e.g. Tegmark et al. 2006) have shown the remarkable consistency
between two-point statistics of the galaxy distribution at $z \sim 0$
and the CMB power spectrum which probes matter clustering at the
recombination.  Mapping the large-scale structure at $z \sim 1$ is thus
crucial to further test the coherency of the gravitational instability
picture at an intermediate time between the epoch of last scattering
and today.

In this section we present a reconstruction of the 3D galaxy density
field, discussing first the methodology and summarising the techniques
adopted to correct for observational selection effects.  These are
fully presented in Marinoni et al. (2005, hereafter Paper I) and
Cucciati et al. (2006), to which the reader is referred for more
details.

\subsection{Density reconstruction method}
\label{method}

The continuous galaxy density fluctuation field

\begin{equation}
\delta_g({\bf r,R})=\frac{\rho({\bf r}, R)-\bar{\rho}}{\bar{\rho}}
\end{equation}

\noindent represents the adimensional excess/deficit of galaxies on a
scale R, at any given comoving position ${\bf r}$ with respect to the
mean density $\bar{\rho}$.  As suggested by Strauss and Willick
(1995) we estimate the smoothed number density of galaxies brighter
than $\mathcal{M}_c$ on a scale $R$, $\rho(r, R, <\mathcal{M}_c)$, by
summing over an appropriately weighted convolution of Dirac-delta
functions with a normalised Gaussian filter F

\begin{equation}
\rho({\bf r}, R, <\mathcal{M}_c)=\sum_i\frac{ \int_{0}^{\infty}\delta^{D}(u-|{\bf \Delta r}_i|/R) 
F(u)du}{S(r_i, \mathcal{M}_c)\Phi(m)\zeta(r_i,m) \Psi(\alpha,\delta)}
\label{denesti}
\end{equation}

\begin{equation}
F(u)=\big(2 \pi R^2 \big)^{-3/2} 
\exp \Big[ -\frac{1}{2} u^2\Big] \,\,\,\, .
\end{equation}

\noindent Here $\Delta {\bf r}=({\bf r_i}-{\bf r})$ is the separation
between galaxy positions and the location ${\bf r}$ where the density field is
evaluated.  We compute the characteristic mean density at position
{\bf r} using equation (\ref{denesti}) by simply averaging the galaxy
distribution in survey slices $r \pm R_s$, with $R_s = 400$\hpcp The
four functions in the denominator of equation \ref{denesti} correct
for various observational characteristics:

- $S(r_i,\mathcal{M}_c)$ is the distance-dependent selection function
of the sample.  This function is identically one when a volume-limited
sample is used.  When the full magnitude-limited survey ($17.5< I<24$
in our case) is used, however, this function corrects for the
progressive radial incompleteness due to the fact at any given
redshift we can only observe galaxies in a varying absolute magnitude
range.  While the PDF of galaxy fluctuations will be derived from
volume-limited samples, in the next section we shall make use of this
function when reconstructing a minimum-variance 3D density map from
the full VVDS survey.

The actual values of $S(r,\mathcal{M}_c)$ are derived using the VVDS
galaxy luminosity function (Ilbert et al. 2005), assuming a minimum
absolute magnitude $\mathcal{M}_c = -15+5 \log h$ and accounting for
its evolution as measured from the VVDS itself. A more detailed
discussion of the derivation of the selection function can be found in
Paper I

- $\Phi(m)$ corrects for the slight bias against bright objects
introduced by the slit positioning tool VMMPS/SPOC (Bottini et
al. 2005).

- $\zeta(r_i,m)$ is the correction for the varying spectroscopic
success rate as a function of the apparent $I_{AB}$ magnitude and of
the distance of the object (see Ilbert et al. 2005).

- $\Psi(\alpha,\delta)$ is the angular selection function correcting
for the uneven spectroscopic sampling of the VVDS on the sky (see Fig
1. of Cucciati et al. 2006).  Its purpose is to make allowance for the
different number of passes done by the VIMOS spectrograph in different
sky regions
(a factor which is anyway maximised in the 4-pass sub-area of the
sample).
\smallskip

\begin{figure*}
\begin{center}
\includegraphics[width=160mm,angle=0]{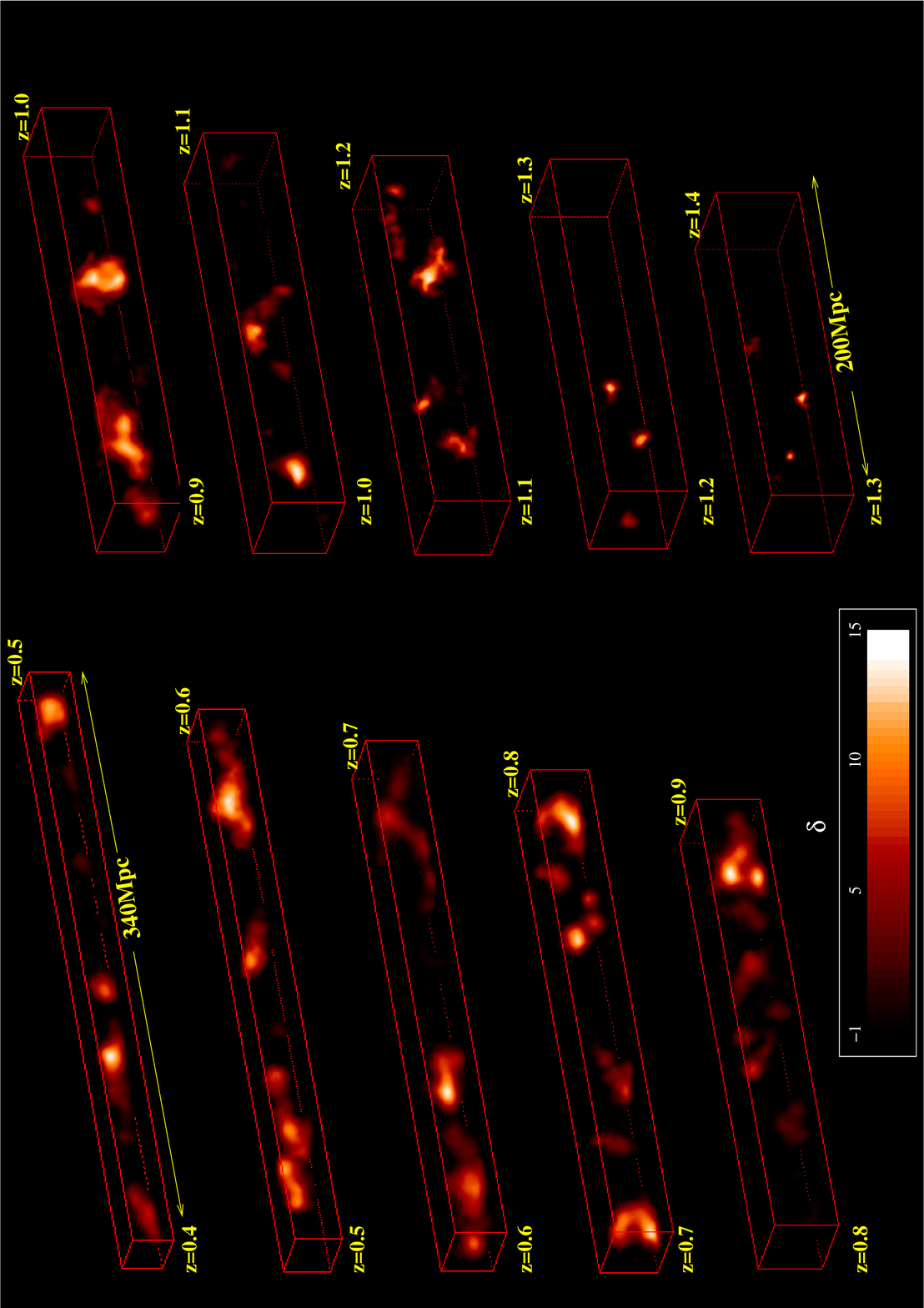}
\caption{ The reconstructed density field for $0.4<z<1.4$, as traced by
  the galaxy distribution in the VVDS-Deep redshift survey to $I \leq
  24$.  This figure preserves the correct aspect ratio between
  transverse and radial dimensions.  The mean inter-galaxy separation
  of this sample at the typical depth of the VVDS ($z=0.75$) is
  $4.6$\hpcv comparable to local redshift surveys as the 2dFGRS.  The
  galaxy density distribution has been smoothed using a 3D Gaussian
  window of radius $R=2$\hpc and noise has been filtered away using a
  Wiener filtering technique (see Strauss \& Willick 1995, Marinoni et
  al. 2005).  Only fluctuations above a signal-to-noise threshold of
  $2$ are shown.  The accuracy and robustness of the reconstruction
  methods have been tested using realistic mock catalogues
  \citep{pol05,mar05}.}
\end{center}
\label{fig1}
\end{figure*}

The analytical form of these selection functions is discussed in
Cucciati et al. (2006). The underlying assumption in this
reconstruction scheme is that the subset of observed galaxies (e.g. in
the case of a flux-limited sample, those luminous enough to enter the
sample at a given redshift) is representative of the full population.
This assumption clearly neglects any dependence of clustering on
luminosity and could bias the density field reconstructed from the pure
flux-limited sample at different redshifts; for this reason, the
quantitative measurements presented in this paper will all be based on
quasi-volume-limited samples, limited to an absolute magnitude
$M_B=-20+5\log h$.  Finally, it should also be mentioned that in
adopting a universal luminosity function we do not take into account
the possible dependence of the luminosity function on morphological
type and environment; this is, however, a second order effect in this
work.

The shot-noise error affecting the reconstructed field at different
{\bf r} is estimated by computing the square root of the variance

\begin{equation} \displaystyle
\epsilon(\blr)=\frac{1}{\overline{\rho_g}}\Bigg[\sum_i
\Bigg(\frac{F\big(\frac{|{\bf \Delta r_i}|}{R}\big)}{S(r_i,\mathcal{M}^c)
\Phi(m_i), \zeta(z,m),\Psi(\alpha,\delta)}\Bigg)^2\Bigg] ^{1/2} \,\, .
\label{shot} \end{equation}

The amplitude of the shot noise increases as a function of redshift in
a purely flux-limited survey.  We deconvolve the signature of this
noise from the density maps by applying the Wiener filter
\citep[cf.][]{pre92,s92b} which provides the {\it minimum variance}
reconstruction of the smoothed density field, given the map of the
noise and the {\it a priori} knowledge of the underlying power spectrum
\citep[\eg][]{lah94}.  For this we assume that the observed galaxy
density field $\delg(\blr)$, and the true (i.e. including all galaxies)
underlying field $\delt(\blr)$, both smoothed on the same scale, are
related via

\begin{equation} \delg(\blr) = \delt(\blr) + \epsilon(\blr),
\end{equation}

\noindent where $\epsilon(\blr)$ is the local contribution from shot
noise (see Eq. (\ref{shot})).  The Wiener filtered density field, in
Fourier space, is

\begin{equation} \tdelf(\blk) = \mathcal{F}(\blk) \tdelg(\blk)\,\, ,
\end{equation}

\noindent where

\begin{equation} \mathcal{F}(\blk) = { \vev{\tdelt^2(\blk)} \over
\vev{\tdelt^2(\blk)} + (2\pi)^{3} P_{\epsilon}(\blk)}\ .
\label{Wiener} \end{equation}

\noindent where brackets denote statistical averages and where
$P_{\epsilon}(\blk)=(2\pi)^{-3} \vev{|\teps^2(\blk)|}$ is the power
spectrum of the noise. Assuming ergodic conditions, this last quantity
can be computed as $P_{\epsilon}(\blk)=(2\pi)^{-3} |\teps(\blk)|^2.$
The calculation of $\vev{\tdelt^2(\blk)}$ taking into account the form
of the window function $F$ and the peculiar VVDS survey geometry is
presented in paper I.

\subsection{A cosmographical tour up to $z=1.5$}
 
We have first applied our reconstruction technique to the global
flux-limited VVDS sample to build a visual three-dimensional map of
galaxy density fluctuations to $z=1.5$ which exploits the full
information content of the survey.  The $I \leq 24$ sample is
characterised by an effective mean inter-particle separation of
($\vev{r} \sim 5.1$ \hpc) in the redshift range [0,1.5]. For
comparison, this sampling is better (denser) than the early CfA1 survey
($\vev{r}\sim 5.5$$ h^{-1}$Mpc) used by \citet{dav81} to reconstruct
the 3D density field of the local Universe (\ie out to $\sim$ 80 \hpc).
Also, at the median depth of our survey, \ie in the redshift interval
$0.7<z<0.8$, the mean inter-particle separation is 4.4 \hpcv a value
nearly equal to the 2dFGRS at its median depth.

The recovered galaxy density field is presented in Fig. 1.
Fluctuations have been smoothed on a scale $R=2$\hpcp  Only density
contrasts with signal-to-noise ratio $S/N>2$ are shown.

A remarkable feature of this {``\em geographical''} exploration of the
Universe at early cosmic epochs is the abundance of large-scale
structures similar in density contrast and size (at least in one
direction) to those observed by local surveys.  In particular, it is
tempting to identify qualitatively a few filament-like density
enhancements bridging more condensed structures along the line of
sight, although the survey transverse size is still too small to fully
sample their extent.  Nevertheless, it is interesting to notice that
these apparently one-dimensional structures remain coherent over scales
$\sim 100$\hpcv separating low-density regions of similar size. Figs.
one and two visually confirm that the familiar web pattern observed in
the local Universe is not a present-day transient phase of the galaxy
spatial organisation but it is already well-defined at $\sim 1.5$ when
the Universe was $\sim 30\%$ its present age \citep[\eg][]{lef96,
  ger05,sco07}.  This implies that large-scale features of the galaxy
distribution essentially reflects the long-wavelength modes of the
initial power spectrum, in agreement with theoretical predictions of
the CDM hierarchical scenario.  Numerical simulations of large scale
structure formation in fact show that the present-day web of filaments
and walls is actually present when the universe was in embryonic form
in the overdensity pattern of the initial fluctuations, with subsequent
linear and non-linear gravitational dynamics just sharpening its
features \citep[\eg][]{bon96, spri05}.

The limited angular size of the survey is exemplified by a dense
``wall'' at $z=0.97$ that stretches across the whole survey solid angle
($0.7 \times 0.7$ deg) (see Fig. 2). This two-dimensional structure is
coherent over more than $\sim 30$\hpc (comoving) in the transverse
direction, is only $\sim 10$\hpc thick along the line of sight, and has
a mean overdensity $\delta_g=2.4 \pm 0.3$.  This makes it similar to
the largest and rarest structures observed in the local Universe, such
as the Shapley concentration \citep[e.g.][]{scar89,bar00}. By applying
a Voronoy-Delaunay cluster finding code \citep{mardav02}, we find 10
distinct groups in this structure, with between 5 and 12 galaxy members
each (down to the limiting magnitude I=24), for a total of 164
galaxies.  If one considers the evolution of {\it mass} fluctuations in
the standard $\Lambda$CDM model, the probability of finding a structure
with similar {\it mass} overdensity at such early times ($0.9<z<1$)
would be nearly 4 times smaller than today: one such {\it mass
  fluctuation} would be expected in a volume of $\sim 3\cdot10^{6}
h^{-3}$Mpc$^3$, \ie nearly 5 times larger than our surveyed volume up
to $z \sim 1$.  In fact, as we shall describe in section 3.3, finding
such a {\it galaxy} overdensity is not so unusual: it is clear evidence
that the {\it biasing} between galaxies and matter at these epochs is
higher than today. This makes fluctuations in the galaxy distribution
to be highly enhanced with respect to those in the mass.

\begin{figure*}
\begin{center}
\includegraphics[width=110mm,angle=-90]{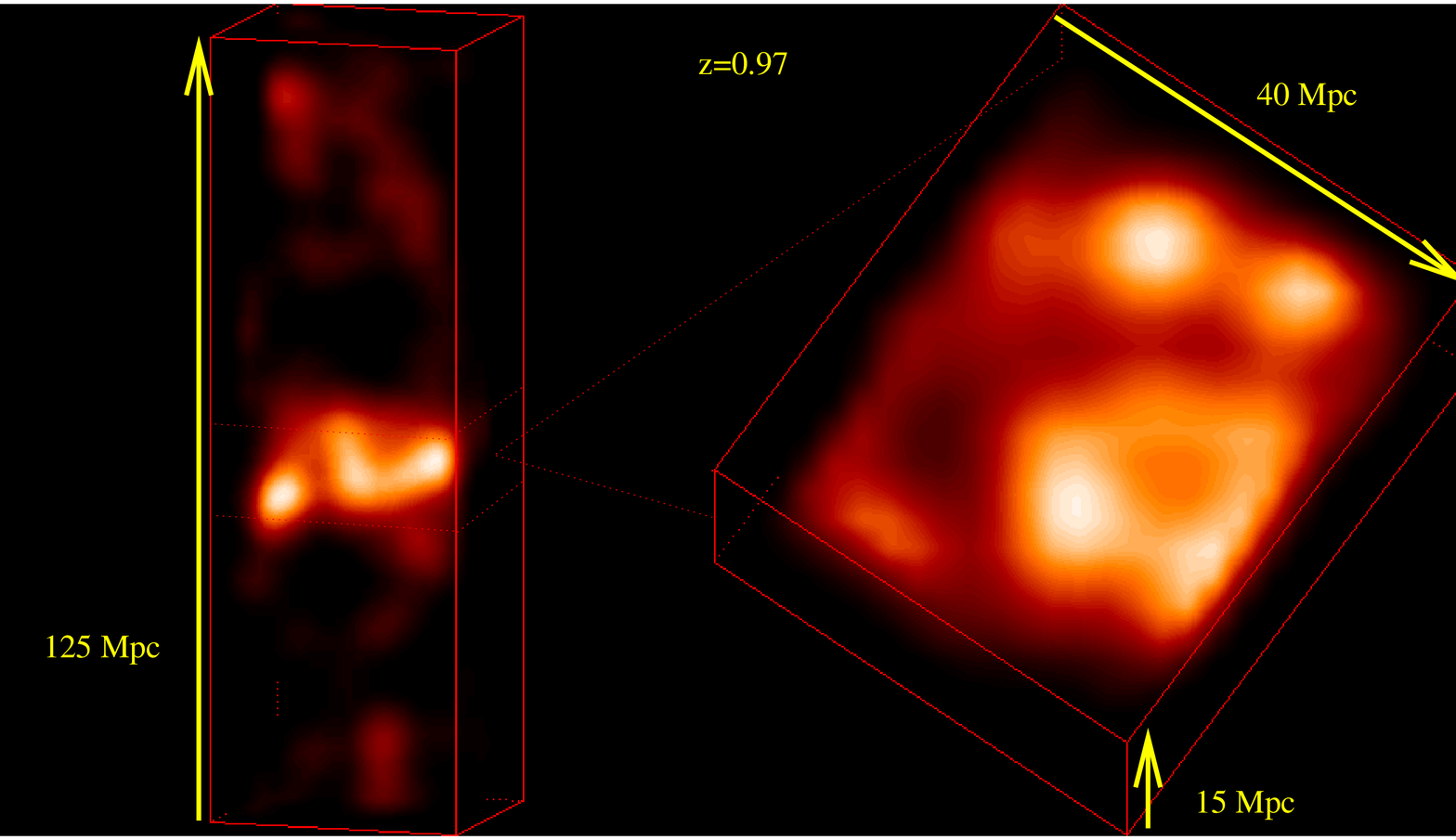}
\caption{
Density distribution and properties of a large-scale planar structure 
at $z=0.97$, that completely fills the VVDS-02h field-of-view.
}
\label{fig2}
\end{center}
\end{figure*}

\subsection{Evolution of the PDFs of galaxy fluctuations in the VVDS}

Several approaches may be used to characterise in a quantitative way
the distribution of galaxy fluctuations $\delg$ shown in Fig 1.  A
complete specification of the overdensity field may be given by the
full set of galaxy N-point correlation functions \citep{dp77}.  This
approach has been explored and routinely applied over the past decade
as better and deeper redshift surveys have become available. An
alternative description may instead be given in terms of the
probability distribution function of a random field. By
definition, the PDF of cosmological density fluctuations describes the
probability of having a fluctuation in the range $(\delta,
\delta+d\delta)$, within a spherical region of characteristic radius R
randomly located in the survey volume.  In principle, it encodes all
the information contained within the full hierarchy of correlation
functions, and provides insights about the time evolution of density
fluctuations.  This definition can be applied either to the
distribution of galaxies, characterizing their number density
fluctuations, or to the dark-matter dominated mass distribution.  For
the latter case, the expected shape of the PDF can be predicted as a
function of redshift given a cosmological model, at least for
large-scale fluctuations; this can be done either analytically (see
below) or using numerical simulations.

On the observational side, in the case of surveys of the local Universe
this fundamental statistics has been often overlooked (but see Marinoni
\& Hudson 2002, Ostriker et al. 2003).  On the other hand, only
recently deep redshift surveys have reached sufficient volumes to allow
these measurements to be extended back in time.  In paper I, we have
discussed and tested in detail the methodology to estimate the PDF from
this kind of samples. In particular we have checked the robustness of
the reconstruction against the specific VVDS-02h survey selection
function, shot-noise errors and other observational biases.  We used
fully realistic mock samples of the VVDS-02h survey data and showed
that, once the smoothing scale $R$ is larger than the mean inter-galaxy
separation, the overall shape of the reconstructed PDF is an unbiased
realisation of the complete parent galaxy population. In particular, we
showed that for redshifts up to $z=1.5$ the VVDS-02h sky coverage and
sampling rate are sufficient for obtaining a reliable reconstruction of
the PDF shape (in both low- and high-density regions) on scales $R \ge
8$\hpcp Clearly, the degree with which the PDF measured from this
sample is a fair representation of the ``universal'' PDF up to \z=1.5
is a separate, yet critical question.  A difference is naturally
expected due to fluctuations on scales larger than the volume probed
(``cosmic variance'').  An estimate of this effect is actually included
in our error bars, as these were drawn from the scatter among our set
of VVDS mock samples.

  We have therefore applied the estimator of Eq. \ref{denesti} and
  the full de-noising technique described in \S~\ref{method} to the
  two luminosity-limited sub-samples of our survey described in
  \S~\ref{data}, reconstructing the PDF of galaxy fluctuations in
  top-hat spheres of radius $R=10$\hpc at two different epochs ($0.7 <
  z < 1.1$ and $1.1 \leq z < 1.5$). 
  The typical luminosity of the galaxies selected in these two
  intervals ($M_{B} \leq -20+5\log h$) corresponds to a
  median luminosity $L_{B}\simeq 2L^{*}_{B}$ at $z\sim 0$
  (i.e. the same median luminosity of the whole 2dFGRS sample \citep{ver02}).
  As discussed
  previously, the use of luminosity-selected samples virtually
  eliminates distance-dependent shot-noise contributions (clearly
  neglecting the residual evolution within the two redshift
  bins, which is well within the errors).  The results are shown in
  Fig. 3.  
  The measured PDF's in Fig. 3 show several interesting features. 
First, as times passes (redshift decreases) the maximum of the PDF
shifts to smaller $\delta$-values; secondly, the low-density tail is
enhanced, with more low-density regions appearing at lower
  redshifts.  Quantitatively, this implies in particular that the
probability of having an under-dense ($\delta_{g}<0$) region of radius
$R=10$\hpc at $0.7 \leq z < 1.1$ is nearly $10\%$ larger than at
earlier times ($1.1 \leq z <1.5$).

\subsection{The expected PDF of mass fluctuations in the mildly
  non-linear regime}

The shape of the PDF of the galaxy overdensities is strongly dependent
on the non-linear effects implicit both in the gravitational growth and
in the physical mechanisms responsible for galaxy formation
\citep[e.g.][]{wat01}.  Initial density fluctuations are normally
assumed to have a Gaussian PDF; this is then modified by the action of
gravity and, in the case of the galaxy field, by the way galaxies trace
the underlying mass ({\em biasing} scheme).  If galaxies were faithful
and unbiased tracers of the underlying mass, the peak shift and the
development of a low-density tail we observe in Fig.3 could be
naturally interpreted as the key signature of dynamical evolution
purely driven by gravity.  In fact, gravitational growth in an
expanding Universe makes low density regions propagate outwards and
become more common as time goes by, while at the same time the
high-density tail increases.

If this interpretation is correct, we expect the PDF of galaxy
overdensities to coincide with the PDF of mass fluctuations
in each redshift range, once they are normalised to the observed
clustering at $z\sim0$, where we know that $L\sim 2L^*$ galaxies 
trace the mass \citep{ver02}.  Let us verify whether this is the case
by first summarising the main formalism to compute the PDF of mass
fluctuations in a given cosmological scenario.

In hierarchical models, it is well established from numerical
simulations that when structure growth reaches the nonlinear regime on
a scale R, the PDF of mass density contrasts in comoving space
is well described by a lognormal distribution
\citep{col91,kof94,tay00,kay01},
\begin{equation}
f_R(\del)=\frac{(2 \pi \omega^2_R)^{-1/2.}}{1+\del} \exp \Big\{ -\frac{
[\ln(1+\del) +\omega^2_R/2]^2}{2\,\omega^2_R} \Big\}\,\,\, . \label{teopdf}
\end{equation}

\noindent This is fully  characterised by a
single parameter $\omega_R$, related to the variance of the
$\del$-field on a scale R as 

\begin{equation} \omega^2_R=\ln [1+\vev{\del^2}_R]
\label{omega}
\end{equation}

At high redshifts, the variance $\sig^2_{R} \equiv \vev{\del^2}_R$ over
sufficiently large scales $R$ (those explored in this paper) is given
in the linear theory approximation by

\begin{equation}
\sig_R(z)=\sig_R(0) D(z) \label{siglt}
\end{equation} 

\noindent where $D(z)$ is the linear growth factor of density
fluctuations (normalised to unity at $z=0$),

\begin{equation}
D(z) = \exp{-\Big[\int_{0}^{z}\rm{f}(z)\,d\ln(1+z)\Big]} \,\,\,\, .
\end{equation}

\noindent In the standard $\Lambda$CDM cosmological model, the
expression for the logarithmic derivative of the growth factor,
$\rm{f}=d\log D/d \log a$ (with $a=(1+z)^{-1}$), can be approximated
to excellent accuracy as
\[\rm{f}(z) \sim \Omega_{m}^{\gamma}(z)\]
\noindent where $\gamma \simeq 0.55$ (Wang \& Steinhardt 1998, Linder
2005) and
\[\Omega_m(z)=\Omega_m^{0}\frac{(1+z)^3}{E^2(z)}\]
\[E^2(z)=\Omega_m^0(1+z)^3+\Omega_{\Lambda}.\]

The lognormal approximation formally describes the distribution of
matter fluctuations computed in real comoving coordinates. On the
contrary, the PDF of galaxies is observationally derived in redshift
space, where its shape is distorted by the effects of peculiar motions
\citep[e.g.][]{mar98,nature07}. In order to map properly the mass
overdensities into galaxy overdensities the mass and galaxy PDFs must
be computed in a common reference frame.  It has been shown by
\citet{sbd00} that an optimal strategy to derive galaxy biasing is to
compare both mass and galaxy density fields directly in redshift space.
Implicit in this approach is the assumption that mass and galaxies are
statistically affected in the same way by gravitational perturbations,
i.e.  that there is no velocity bias in the motion of the two
components.

The relation between the variances measured in real and redshift
comoving space is

\begin{equation} \sig^z_R(z)=p(z)\sig_R(z)
\label{sigcor} \end{equation}

\noindent where $p(z)$ is a redshift-dependent
correcting factor which takes into account the average contribution of
the linear redshift distortions induced by peculiar velocities
\citep{kai87}. Its expression, in the high redshift regime, is given
by \citep{hamilton,mar05}

\begin{equation}
p(z)=\big[1+\frac{2}{3}\rm{f}(z)+\frac{1}{5}\rm{f}^{2}(z)\big]^{1/2}.
\end{equation}

We have used this formalism to compute the PDF expected for mass
fluctuations in the redshift ranges explored using our galaxy samples.
This is given, for the two ranges, by the curves in the top panel of
Fig. 3.  The evident discrepancy between the galaxy and mass PDF's
indicates that the observed evolution cannot be only the product of
gravitational growth (in the adopted cosmological model), but that a
time-evolving bias between the galaxy and mass density fields is
needed: at high redshifts and on large scales galaxy overdensities
trace in a more biased way the underlying pattern of dark matter
fluctuations. In the following section we shall summarise our current
knowledge on the properties and evolution of the biasing function and
show how the presence of a non-linear bias is a necessary ingredient to
theoretically understand the evolution of the PDF in Fig.3. This will
in particular provide us with the necessary background to interpret the
evolution of the low-order moments of the PDF at different redshifts,
which is our aim in this paper.

\begin{figure}
\includegraphics[width=74mm,height=71mm, angle=0]{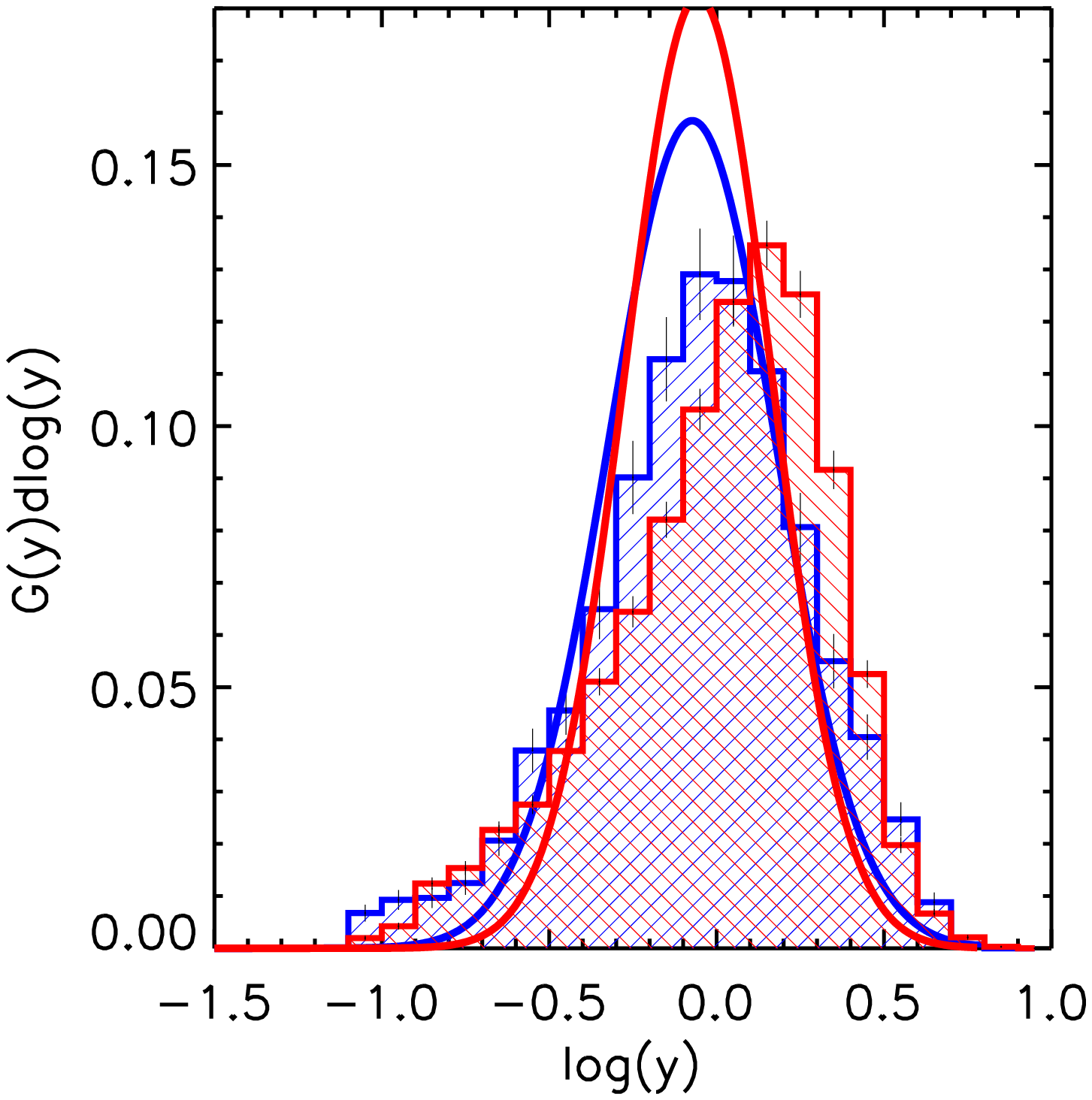}
\includegraphics[width=74mm,height=71mm, angle=0]{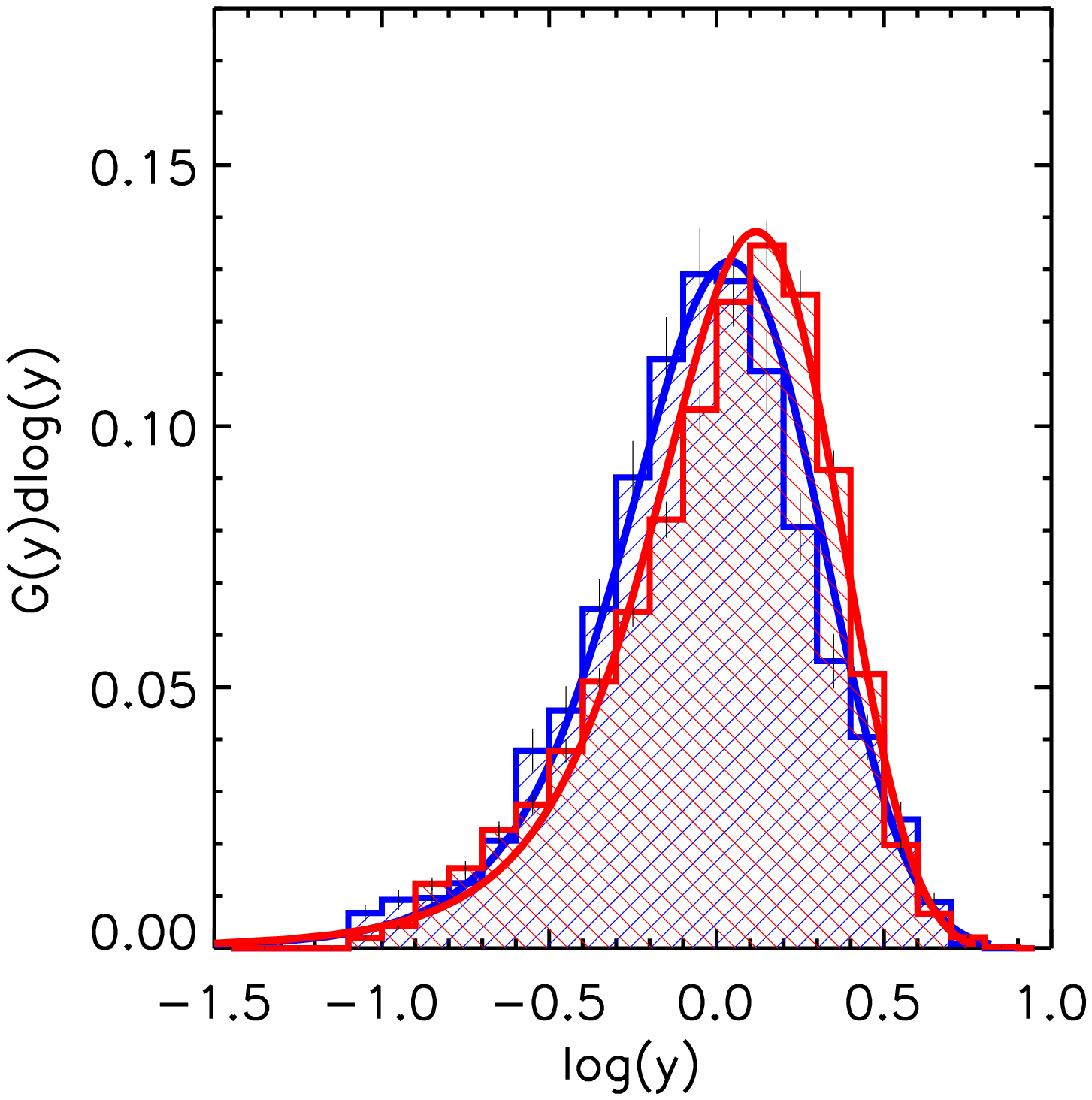}
  \caption{ The PDF of galaxy fluctuations (in units $y=1+\delta$) for
    VVDS galaxies with $M_{B} \leq -20+5 \log h$ from the VVDS within
    two independent volumes, corresponding to different cosmic epochs:
    $0.7 < z <1.1$ (blue shaded histogram), and $1.1 \leq z <1.5$
    (green shaded histogram). The galaxy PDF has been reconstructed using a
    Top-Hat smoothing window of comoving size $R=10$\hpcp  The
    histograms actually correspond to the distribution function $G(y)
    = \ln(10) y g(y)$ because the binning is done in $\log(y)$.  
The two observed histograms have been reproduced in both the upper and
lower panels.  They are compared to the theoretical predictions 
for the PDF of, respectively, 
mass fluctuations (top, from Eq. \ref{teopdf}) 
and of {\em galaxy} fluctuations as inferred from Eq. \ref{fife} using  the 
    non-linear biasing function measured from the VVDS (bottom). 
    The blue and
    red lines correspond to the higher- and lower-redshift samples
    respectively. }
\label{fig3}
\end{figure}

\subsection{Evolution and non-linearity of biasing}

Biasing lies at the heart of all interpretations of large-scale
structure models.  Structure formation theories predict the
distribution of mass; thus, the role of biasing is pivotal in mapping
the observed light distribution back into the theoretical model.  In
our case we need to disentangle the imprint of biasing from that of
pure gravity in the evolution of the galaxy PDF.

In paper I we inferred the biasing relation
$\delg=\delg(\del)=b(\delta)\delta$ between mass and galaxy
overdensities from their respective probability distribution functions
$f(\del)$ and $g(\delg)$. Assuming a one-to-one mapping between mass
and galaxy overdensity fields, conservation of probabilities
implies \citep[e.g.][]{sbd00,wild}
\begin{equation}
g(\delta_g)d\delta_g=f(\delta)d\delta
\label{fife}
\end{equation}

This approach implies the assumption of a cosmological model (the
standard $\Lambda$CDM model in our case) from which to compute
$f(\delta)$, the mass PDF.  The advantage over other methods is that we
can explore the functional form of the relationship $\delg=b(z,\del, R)
\del$ over a wide range in mass density contrasts, redshift intervals
and smoothing scales R without imposing any {\it a-priori} parametric
functional form for the biasing function.  Note that, by definition,
this scheme is ineffective in capturing information about possible
stochastic properties of the biasing function.

The numerical solution $\delta_g=\delta_g(\delta)$ of Eq. \ref{fife}
maps the mass PDF (solid lines in the top panel of Fig. 3) into the
galaxy PDF (solid lines in the bottom panel of Fig. 3) and can be
analytically approximated using a Taylor expansion \citep{fg93}
\begin{equation}
\delta_g(\delta)=\sum_{k=0}^{n}\frac{b_{k}(z)}{k!}\delta^{k},  
\label{nlbf}
\end{equation} 
\noindent where the coefficients $b_i$ depend on redshift.  We consider
this power series only to second order, and fit the numerical solution
for the biasing function leaving $b_0$ as a free parameter.  Avoiding
setting $b_0$ as an integral constraint ($<\delta_g>=0$) allows us to
account for possible (un-modelled) contributions from higher order
moments of the expansion. This approach has the advantage of minimising
biases in the estimates of the lower moments of the expansion,
specifically $b_1$ and $b_2$.

The key result from Paper I has been to show that galaxy biasing is
poorly described in terms of a single scalar and better characterised
by a more sophisticated representation.  Specifically, always
considering a scale $R=10$\hpcv the ratio between the quadratic and
linear bias terms has been evaluated in four different high redshift
intervals (see Table 2 of Paper I).  When averaged over the full
redshift baseline $0.7<z<1.5$, this ratio turns out to be
\begin{equation}
\left\langle \frac{b_2}{b_1} \right\rangle=-0.19 \pm 0.04 \label{rationl}
\end{equation}
\noindent i.e. different from zero at more than 4$\sigma$ confidence
level.  This means that -- at least over the redshift range and scales
considered here -- the level of biasing depends on the underlying value
of the mass density field. In other words, the way galaxies are
distributed in space depends in a non-linear manner on the local
amplitude of dark matter fluctuations.

The measurement of a non-linear term in the biasing relation is fully
consistent with a parallel analysis of the hierarchical scaling of the
N-point correlation functions in the same VVDS sample (Cappi et al.
2008).  These results confirm a generic prediction of hierarchical
models of galaxy formation \citep[\eg]{som01}.  It is relevant to
compare them to estimates of the bias function at the current epoch.
Early works indirectly suggested that also at $z\sim 0$ the biasing
function should have a non-negligible non-linear component.  Comparing
the two--point correlation functions and the normalized skewness of
SSRS2 galaxies, Benoist et al. (1999) showed that the relative bias
between galaxies with different luminosity is non-linear, which
indirectly indicates that (at least for luminous galaxies) the bias
with respect to the dark matter must be non-linear as well. A similar
analysis was performed by Baugh et al. (2004) and Croton et al. (2004)
on the 2dFGRS, finding results consistent with Benoist et al.; finally,
Feldman et al. (2001) and Gazta\^{n}aga et al. (2005) directly measured
the three--point correlation function (in both Fourier and real space)
for the IRAS and 2dFGRS samples respectively and found evidence for
$b_2 < 0$.
  
On the other hand, these results seem to be inconsistent with another
analysis of the 2dFGRS performed using the bi-spectrum \citep{ver02}.
Hikage et al. 2005 by analysing the SDSS galaxies with a Fourier-phase
technique also conclude that the bias in this survey is essentially
linear.  If one ignored the other independent analysis of the 2dFGRS,
it could be speculated that the large-scale non-linear term that we
detect at $z>0.7$ is suppressed as a function of cosmic time; this is
however not supported by the results of numerical experiments
\citep{som01}.  Interestingly, we note that when compared to the local
non-linear measurements ($b_2/b_1\sim -0.35$) \citep{fel01,gaz05}, our
estimate seems to suggest that the amplitude of the quadratic term
$b_2/b_1$ decreases (in absolute terms) as a function of redshift, a
results in qualitative agreement with indications from simulations.  It
seems therefore more likely that the difference in the reconstruction
methods used (with different sensitivity to higher order terms in Eq.
\ref{nlbf}) is a better explanation of the discrepant results at $z\sim
0$. We will show in the next section (\S 4) how the self-consistency of
the evolution of the variance and skewness of galaxy counts with
redshift, indeed requires the presence of a non-linear biasing
component.

The information contained in the non-linear function of Eq. \ref{nlbf}
can be compressed into a single scalar term that can be used to
interpret the evolution of two-point statistics (correlation function)
as well as the variance of the galaxy density field (see \S 4.2).
Since, by definition, $\vev{b(\del) \del}=0$, the most interesting
linear bias estimators are associated to the second order moments of
the PDFs, \ie the variance $\vev{\delg^2}$ and the covariance
$\vev{\delg\,\del}$.  Following the prescriptions of \citet{del99}, we
characterize the biasing function as follows

\begin{equation} 
\bl^2 \equiv \frac{\vev{b^2(\del)\, \del^2}}{\vev{\del^2}} 
\label{linb}
\end{equation}

\noindent where $\bl^2$ is an estimator of the linear biasing parameter
defined, in terms of the two-point correlation function, as
$\xi_g=b_L^2 \xi$. We evaluate Eq. \ref{linb} using Eq. \ref{nlbf} with
parameters $b_i(z)$ estimated locally by Verde et al. (2002) and in the
redshift range $0.4<z<1.5$ by Marinoni et al. (2005).  The best fitting
phenomenological model describing the redshift scaling of the linear
biasing parameter for a volume limited population of ``normal''
galaxies with median luminosity $L \sim 2L^{*}(z=0)$ is
\[b_{L}(z)=1+(0.03\pm0.01) (1+z)^{3.3\pm0.6}\]

While today $\sim 2L^{*}$ galaxies trace the underlying mass
distribution on large scales \citep{lah02, ver02, gaz05}, in the past
the two fields were progressively dissimilar and the relative biasing
systematically higher. In Paper I we showed how this observed redshift
trend compares to different theoretical models for biasing evolution,
i.e.  a ``galaxy conserving'' model (Fry et al. 1996), a ``halo
merging'' model (Mo \& White 1996) and a ``star forming'' model
(Tegmark \& Peebles 1998).

\section{Testing gravitational instability 
with the low-order moments of the PDF}

Having decoupled biasing effects from the purely gravitational
evolution of the galaxy PDF we have now all the ingredients to use this
latter quantity to test the consistency of some general predictions of
the GIP.  The evolution of the low-order statistical moments of the
galaxy PDF, specifically its second and third moments can be compared,
on large scales with analytical predictions of linear and second order
perturbation theory respectively.

\subsection{Estimating the moments from redshift survey data}

Following standard conventions, we define the second- and third-order
moments, on a scale $R$, of a continuous, zero-mean overdensity field
as

\begin{equation} \vev{\delg^2}_{R} =
\int_{-1}^{\infty} \delg^2 g_{R}(\delg) d \delg.  \label{var}
\end{equation}
\noindent and
\begin{equation} 
\vev{\delg^3}_{R}=
\int_{-1}^{\infty} \delg^3 g_{R}(\delg) d \delg.  \label{skew}
\end{equation}

Note that the moments cannot be estimated as ensemble averages over the
reconstructed PDF. In fact, this last quantity has been reconstructed
using the Wiener filtering technique.  This minimises the shot noise
contribution (\S~\ref{method}) but gives a biased estimate of the
density field moments (via Eq. \ref{var} and \ref{skew}) as it requires
an input power spectrum (and therefore {\it assuming} a second moment).
A standard practical way to estimate moments is to randomly throw
spherical cells down within the galaxy distribution and reconstruct the
count probability distribution function $P_k=n_k/N$ (where $n_k$ is the
number of cells that contain k galaxies out of a total number of cells
N. The moments are then estimated as
\begin{equation}
\vev{\delg^p}=\bar{N}^{-p} \sum_{k=0}^{\infty} P_k
\vev{(k-\bar{N})^p} 
\end{equation}
where $\bar{N}=\sum_{k=0}^{\infty}k P_k$.

The quantities we are interested in are the cumulants
$\vev{\delta^{p}}_c$ of the one-point density PDF. For a density field
smoothed with a top-hat window, the $p$-order cumulant
\begin{equation} 
\vev{\delg^{p}}_c=\frac{1}{v^{p}_{R}}\int \xi_{p}(r_1,r_2...r_p)d^3r_1d^3r_2...d^3r_p
\end{equation}

is the average of the N-point reduced correlation function over
the corresponding cell of volume $v_R$ (from now on we will only
consider the scale $R=10$\hpc and we will drop the suffix $R$, unless
we need to emphasize it). This is defined as the connected part of the
N-point correlation function $\vev{\delg(r_1)\delg(r_2)...\delg(r_p)}$
in such a way that for $p>2$ $\xi_p=0$ for a Gaussian field. Since
the galaxy distribution is a discrete process (Eq.~\ref{denesti} is
a sum over Dirac delta functions) and since, by definition, the
density contrast has a zero mean, the connection between low-order
cumulants and moments is given by

\begin{equation}
\vev{\delg^2}_c=\vev{\delg^2}-\frac{1}{\bar{N}} 
\label{varp}
\end{equation}

\begin{equation}
\vev{\delg^3}_c=\vev{\delg^3}-3\frac{\vev{\delg^2}_c}{\bar{N}}-\frac{1}{\bar{N}^2}
\end{equation}

These relations accounts for discreteness effects using the Poisson
shot-noise model \citep[\eg][]{pee80,fry85}. Possible biases introduced
by this technique are discussed by Hui \& Gazta\~naga 1999, while an
alternative approach is detailed by Kim \& Strauss (1998).

Finally, it is necessary to devise a strategy to compensate for the
fact that a cell will sample regions that have varying angular and
spectroscopic completeness and which may even span the survey boundary.
For this reason the galaxy counts are scaled up in proportion to the
degree of incompleteness in the cell. This is done by weighting galaxy
counts using the selection functions $\Phi(m)$, $\zeta(r, m)$, and
$\Psi(\alpha,\delta)$ defined in \S~\ref{method}.  Additionally,
although our reconstruction scheme accounts for the non-uniform VVDS
angular coverage, we further restrict ourselves to counts in spheres
having at least 70\% of their volume in the denser 4-pass region, in
order to avoid possible edge effects.  We remark that moments are
estimated from virtually volume-limited samples, as defined in
\S~\ref{data}.  As a consequence, the radial selection function is
constant and any variations in the density of galaxies are due only to
large-scale structure.

\subsection{The evolution of the {\it rms} and skewness of 
  galaxy fluctuations}

Since in perturbation theory higher order cumulants are predicted to
be a function of the variance, it is useful, in the following, to
define the normalized skewness 
\begin{equation}
S_3=\vev{\delg^3}_c/\sigma^2 \,\, ,
\end{equation}

where the shot-noise corrected variance $\sigma^2$ is given by Eq.
\ref{varp}.  Fig. 4 shows the evolution of the {\it rms} fluctuation
and the normalized skewness on a scale $R=10$\hpcv as measured from the
VVDS volume-limited sub-samples.  Errors have been computed using the
50 fully-realistic mock catalogs of VVDS-Deep discussed in Pollo et al.
(2005).  This allows us to include an estimate of the contribution of
cosmic variance, which represents the most significant term in our
error budget.

The top panel of Fig.~4 shows that the square-root of the
variance, which measures the r.m.s. amplitude of fluctuations in
galaxy counts, is with good approximation constant over the full
redshift baseline investigated: in redshift space, the mean value of
$\sigma_g$ for our volume-limited galaxy samples 
is $0.78\pm 0.09$ for $0.7<z<1.5$.
A similar, nearly constant value is also consistent with the
  value estimated at $z\sim 0.15$ from the 2dF galaxy redshift survey
 \citep{cro04} that is also reported in same figure. 
  This means that over nearly 2/3 of the age of the Universe the
observed fluctuations in the galaxy distribution look almost as
frozen, despite the underlying gravitational growth of mass
fluctuations. This quantifies the visual impression we had from Fig.~1,
  that the distribution of galaxies is as inhomogeneous
  at $z\sim 1$ as it is today.

\begin{figure}
\includegraphics[width=90mm,angle=0]{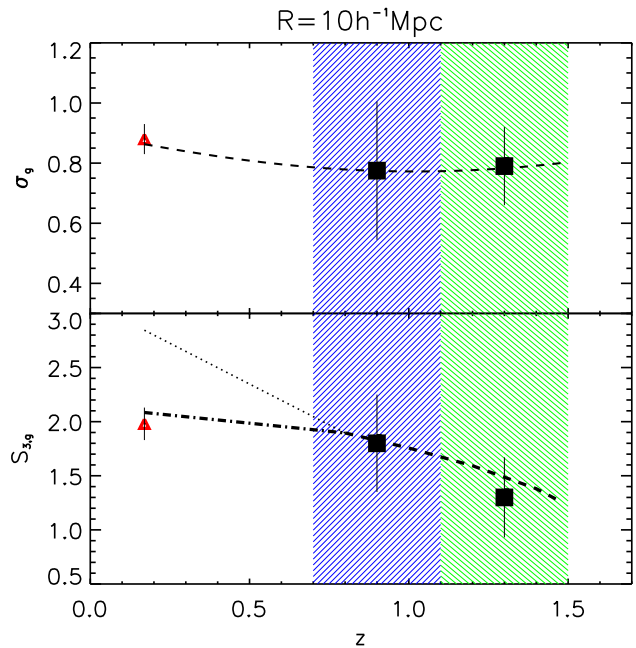}
\caption{Evolution of the {\it r.m.s} deviation (top) and skewness
  (bottom) of the PDF of galaxy fluctuations on a scale $R=10$\hpcp The
  filled squares correspond to two volume-limited samples from the VVDS
  with $M_{B}< -20+5\log h$ covering the redshift intervals indicated
  by the shaded regions.  Triangles correspond to the 2dFGRS
  measurements at $z\simeq 0.15$ \citep{cro04}, from a sample including
  similarly bright galaxies.  Error bars give 68\% confidence errors,
  and, in the case of VVDS measurements, include the contribution from
  cosmic variance. The dashed lines in both panels show the theoretical
  predictions for the evolution of the variance (Eq. \ref{var1}) and
  skewness (Eq. \ref{skew2}) inferred using VVDS measurement of
  biasing.  Predictions for the skewness (based on the
  $(b_1(z),b_2(z))$ measurements in the redshift range $0.7<z<1.5$
  quoted in Table 2 of Paper I) have been extrapolated to $z\sim 0$
  using the local (2dFGRS) biasing measurements of Verde et al. (2002)
  (linear bias, dotted line) and of Gazta\~{n}aga et al. (2005)
  (quadratic bias with $b_2/b_1=-0.34$, dot-dashed line).  }
\label{fig5}
\end{figure}

The third moment, which measures asymmetries between under- and
over-dense regions, indicates that the galaxy density field was
non-Gaussian on large scales (10 \hpc) even at these remote epochs
($\sim 4 \sigma$ detection).  In particular we find indication for an
increase of the normalised skewness with cosmic time, when comparing
the VVDS values to the local measurement by 2dFGRS.

Using the measured bias evolution, we can translate the specific
predictions of the GIP for the variance and skewness of the matter
density field into the corresponding observed quantities. Using linear
perturbation theory, the scaling of the {\it rms} of number density
fluctuations is

\begin{equation}
\sigma_{g}(z) \sim b_{L}(z)D(z)p(z)\sigma(0) \,\,\, ,
\label{var1}
\label{mia}
\end{equation}

In a Universe in which primordial density fluctuations were Gaussian,
the non-linear nature of gravitational dynamics leads to the emergence
of a non-trivial skewness of the local density PDF.  

Within the
framework of gravitational perturbation theory, the first non-vanishing
term describing the evolution of the skewness of a top-hat filtered,
initially Gaussian matter density field corresponds to second-order.

According to non-linear, second-order perturbation theory predictions,
the skewness of the mass distribution is approximately independent of
time, scale, density, or geometry of the cosmological model. Assuming
that its evolution only depends on the hypothesis that the initial
fluctuations are small and quasi-Gaussian and that they grow via
gravitational clustering one derives that, in redshift-distorted space
\citep{pee80, jbc93, ber94, hiv95}

\begin{equation}
S_{3}\sim \frac{35.2}{7}-1.15(n+3)
\end{equation}

\noindent where $n$ is the effective slope of the power spectrum on the
scales of interest [i.e. in our case, since $R=10$\hpc, n is
approximately given by -1.2 \citep[e.g.][]{ber02}].  Subsituting the
evolution of bias in second order approximation, the evolution of the
observed skewness is given by \citep{fg93}

\begin{equation}
S_{3,g}\sim b_{1}(z)^{-1} \Big[S_3 + 3 \frac{b_2(z)}{b_1(z)} \Big]. 
\label{skew2}
\end{equation}

The curves in both panels of Fig. 4 show that equations (\ref{var1})
and (\ref{skew2}) reproduce extremely well the evolution of variance
and skewness observed within the VVDS.

The mass PDF is a one-parameter family of curves completely specified
once the linear evolution model for the mass variance $\vev{\del^2}$ is
supplied. This implies that our {\it non-linear} biasing estimate is
fully independent from predictions of higher-order perturbation theory.
On the contrary, non-linear biasing at $z=0$ is inferred by directly
matching 3-point galaxy statistics with the corresponding mass
statistics derived from weakly non-linear perturbation theory (e.g.
Verde et al. 2000, Gazta\~{n}aga et al. 2005).  As a consequence, the
agreement we find between predicted and observed third-order moments is
not a straightforward consequence of the method used to derive the
biasing function.  These results provide an indication of the
consistency, at $z=1$, of some constitutive elements of the standard
picture of gravitational instability from Gaussian initial conditions.

Concerning the local measurements from
2dFGRS, the predicted scaling for the skewness continues to show very good
agreement if the local, non-linear measurement of Gazta\~{n}aga et
al. (2005) is considered. The value of $S_{3,g}$, however,
cannot be consistent with GIP predictions if in the local
universe the simple linear biasing measurement of Verde et
al. (2002) (i.e. $b_2=0$) is adopted.

\subsection{Effect of galaxy luminosity evolution}

In the above comparison of galaxy samples at three different epochs, we
have so far neglected an important point.  Galaxy luminosity evolves
significantly between $z\sim 0$ and $z \sim 1$, with a mean brightening
of at least $ 1$ magnitude for an average spectral type (Ilbert et al.
2005).  Thus, the contribution to the clustering signal at
progressively earlier epochs may not be be due to the progenitors of
the galaxies that are sampled at later times in the same luminosity
interval. Luminosity evolution between $z=1$ and $z=1.5$ is more
uncertain but certainly smaller due to the shorter time interval.  The
brightening between the two VVDS sub-samples considered is expected not
to be very significant for these very luminous objects.

To compare galaxies at $z=1$ and $z\sim 0$ one should in principle
confront our high-redshift results with those of local galaxies about
one magnitude fainter.  According to 2dFGRS results, this means
shifting the local estimates of variance to a slightly smaller value
(Norberg et al. 2002) and leaving the skewness measurement unchanged
within the quoted error bars (Croton et al. 2004).  These changes would
only reinforce our conclusions about the evolution of the low order
moments of the PDF of galaxy fluctuations. In particular, since a
fainter sample has a smaller bias threshold, the locally measured
skewness would make the discrepancy with GIP predictions for a simple
linear biasing model even stronger.

\section{Conclusions}

The results presented in this paper provide the first direct evidence
at $z\sim1$ for the consistency of the GIP hypothesis as described in the
framework of general relativity.  The standard theory of structure
formation via gravitational instability successfully explains the
present day statistics (e.g. Tegmark et al. 2006) and dynamics (e.g.
Peacock et al. 2001) of large scale structures. We have shown that
observations are fully consistent with these predictions over the
entire redshift baseline $0<z<1.5$ once the biasing between the galaxy
and matter distribution is properly described. In Paper I we showed
that it is necessary to include a small (10\%) yet crucial non-linear
component to accurately account for the observed probability
distribution function of galaxy overdensities.  Here we have shown that
this component is also necessary as to understand the observed
evolution of the low-order moments of the galaxy overdensity field.

More specifically, our analysis of the 3D density fluctuation field
traced by a volume-limited sample of VVDS galaxies (with $M_{B} \leq
-20 +5 \log h$) at different epochs unambiguously reveals the
time-dependent effects of gravitational evolution:

\indent a) underdense regions progressively occupy a larger volume
fraction as a function of cosmic time, as expected from gravitational
growth in an expanding background;

\indent b) the second moment of the field traced by this ``normal''
population of galaxies (with median luminosity $\sim 2L*$) is
statistically consistent with the local ($z \sim 0$) estimate for
similarly luminous galaxies, i.e. it is approximately constant over the
full redshift baseline $0<z<1.5$.  This implies that the apparent
inhomogeneity in the galaxy distribution remains similar, i.e.
galactic fluctuations have almost frozen over nearly 2/3 of the age of
the universe \citep{giav98,coil04,pol06}.  We have shown that this is
readily explained by the combination of the gravitational growth of
mass fluctuations with the evolution of the bias between galaxies and
mass. These two factors almost cancel each other out;

\indent c) there are some hints that the skewness increases with cosmic
time, its value at $z\sim 1.5$ being nearly 2$\sigma$ times lower than
that measured locally by the 2dFGRS for similarly luminous galaxies.
In particular, the measured value of the skewness at $z\sim 1.5$ (on
scales $R=10$\hpc) indicates that galaxy fluctuations are strongly
non-Gaussian ($\sim 4 \sigma$ detection) even at such an early epoch
(see Cappi et al. 2008 for a different approach which arrives at
similar conclusions);

\indent d) remarkably, once VVDS measurements of non-linear biasing are
included, both these trends are consistent with predictions of linear
and second-order perturbation theory for the evolution of gravitational
perturbations as described within the framework of general relativity;

\indent e) we have shown that the values of the skewness we measure at
high redshift are difficult to reconcile with the 2dFGRS measurements
if local biasing is linear \citep{ver02}. A fully coherent
gravitational picture emerges, however, over the whole baseline
$0<z<1.5$ if the non-linearity of the local biasing function is taken
into account, at the level estimated by Feldman et al. 2001 and
Gazta\^{n}aga et al. 2005. Compared to these local measurements our
results seem to suggest that the amplitude of the quadratic term
$|b_2/b_1|$ is a decreasing function of redshift at least up to $z\sim1.5$.

\section*{Acknowledgments}
We thank the referee, M. Strauss, for important suggestions that
significantly improved the manuscript. LG acknowledges the hospitality of
MPE and the ``Excellence Cluster Universe'' in Garching, where part of
this work was completed. 
This research has been developed within the framework of the VVDS
consortium and it has been partially supported by the CNRS-INSU and
its Programme National de Cosmologie (France), and by the Italian
Ministry (MIUR) grants COFIN2000 (MM02037133) and COFIN2003
(num.2003020150).  The VLT-VIMOS observations have been carried out on
guaranteed time (GTO) allocated by the European Southern Observatory
(ESO) to the VIRMOS consortium, under a contractual agreement between
the Centre National de la Recherche Scientifique of France, heading a
consortium of French and Italian institutes, and ESO, to design,
manufacture and test the VIMOS instrument.

\label{lastpage}


\begin{thebibliography}{99}


\bibitem[\protect\citeauthoryear{Bardelli et al.}{2000}]{bar00}
Bardelli, S. et al. 2000, MNRAS, 312, 540 

\bibitem[\protect\citeauthoryear{Baugh et al.}{2004}]{bau04}
Baugh C.M. et al., 2004, \mnras, 351, L44

\bibitem[\protect\citeauthoryear{Benoist et al.}{1999}]{benoista}
Benoist, C., Cappi, A., da Costa, L.N., 
Maurogordato, S., Bouchet, F.R., Schaeffer, R., 1999, \apj, 514, 563

\bibitem[\protect\citeauthoryear{Bernardeau}{1994}]{ber94}
Bernardeau, F. 1994,  ApJ,  433, 1 

\bibitem[\protect\citeauthoryear{Bernardeau et al.}{2002}]{ber02}
Bernardeau, F., Colombi S., Gazta\~{n}aga E., \& Scoccimarro, R. 2002,
Phys. Rep. 367, 1


\bibitem[\protect\citeauthoryear{Bond et al.}{1996}]{bon96}
Bond, J. R.,  Kofman, L., \& Pogosyan, D., 1996, Nature, 380, 603 

\bibitem[\protect\citeauthoryear{Bouchet et al.}{1995}]{bou95}
Bouchet F. R., Colombi S., Hivon E., \& Juszkiewicz R. 1995, A\&A, 296, 575

\bibitem[\protect\citeauthoryear{Cappi et al.}{2008}]{cap08}
Cappi, A. et al. (the VVDS team) 2008, in preparation

\bibitem[\protect\citeauthoryear{Coil et al.}{2004}]{coil04}
Coil, A. F. et al., 2004, \apj, 609, 525

\bibitem[\protect\citeauthoryear{Coles \& Jones}{1991}]{col91} 
Coles, P., \& Jones, B. 1991, \mnras, 248, 1

\bibitem[\protect\citeauthoryear{Croton et al.}{2004}]{cro04}
Croton, D. J., et al. 2004, MNRAS, 352, 1232 

\bibitem[\protect\citeauthoryear{cucciati et al.}{2006}]{cuc06} 
Cucciati, O., et al. (the VVDS team) 2006, A\&A, 458, 39

\bibitem[\protect\citeauthoryear{da Costa et al.}{1994}]{dac94} 
da Costa, L.N., et al. 1994, ApJ 424, L1

\bibitem[\protect\citeauthoryear{Davis \& Peebles}{1977}]{dp77}
Davis, M., \& Peebles, P. J. E. 1977, ApJS, 34, 425

\bibitem[\protect\citeauthoryear{Davis \& Huchra}{1981}]{dav81}
Davis, M., \& Huchra, J. 1981, \apj, 254, 437

\bibitem[\protect\citeauthoryear{Davis et al.}{2003}]{dav03}
Davis, M. et al. (the DEEP2 team)  2003, Proc. SPIE, 4834, 161

\bibitem[\protect\citeauthoryear{Dekel \& Lahav}{1999}]{del99}
Dekel, A., \& Lahav, O. 1999, \apj, 520, 24


\bibitem[\protect\citeauthoryear{Feldman et al.}{2001}]{fel01}
Feldman, H. A., Frieman, J. A., Fry, J. N., Scoccimarro, R., 2001, Phys.Rev.Lett. 86, 1434

\bibitem[\protect\citeauthoryear{Fry}{1985}]{fry85} Fry, J. N.  1985, \apj, 289, 10

\bibitem[\protect\citeauthoryear{Fry \& Gazta\~{n}aga}{1993}]{fg93}
Fry, J. N., \& Gazta\~{n}aga, E.  1993, ApJ,  413, 447 

\bibitem[\protect\citeauthoryear{Gazta\~{n}aga et al.}{2005}]{gaz05}
Gazta\~{n}aga, E.  et al. 2005,  	 2005, \mnras, 364, 620

\bibitem[\protect\citeauthoryear{Geller \& Huchra}{1991}]{gh89}
Geller, M, J., \& Huchra, J.P. 1989, Science, 246, 897

\bibitem[\protect\citeauthoryear{Gerke et al.}{2005}]{ger05}
Gerke, B. F.  et al. 2005, \apj, 625, 6

\bibitem[\protect\citeauthoryear{Giavalisco et al.}{1998}]{giav98}
Giavalisco, M. et al. 1998, ApJ, 503, 543


\bibitem[\protect\citeauthoryear{Giovanelli \& Haynes}{1991}]{gh91}
Giovanelli, R., \& Haynes, M. 1991,  ARA\&A,  29, 499 

\bibitem[\protect\citeauthoryear{Guzzo et al.}{2008}]{nature07}
Guzzo, L. et al. (the VVDS team) 2008, Nature, 451, 541

\bibitem[\protect\citeauthoryear{Hamilton}{1998}]{hamilton}
Hamilton A.J.S., 1998, in ``The Evolving Universe'', Kluwer Astrophysics and
Space Science Library, 185, 231,  (also astro-ph/9708102).

\bibitem[\protect\citeauthoryear{Hivon et al.}{1995}]{hiv95}
Hivon, R., Bouchet, F. R., Colombi, S., \& Juszkiewicz, R. A\&A, 298, 643

\bibitem[\protect\citeauthoryear{Hui \& Gazta\~naga}{1999}]{hui99}
Hui, L., Gazta\~naga, E.  1999, \apj, 519, 622

\bibitem[\protect\citeauthoryear{Ilbert et al.}{2005}]{ilb05}
Ilbert, O. et al. (the VVDS team) 2005, A\&A, 439, 863

\bibitem[\protect\citeauthoryear{Juszkiewicz, Bouchet, \& Colombi}{1993}]{jbc93}
Juszkiewicz, R., Bouchet, F. R., \& Colombi, S. 1993, ApJ, 412, L9 (34)

\bibitem[\protect\citeauthoryear{Kaiser}{1984}]{kai84}
 Kaiser, N. 1984, ApJ, 284L, 9 

\bibitem[\protect\citeauthoryear{Kaiser}{1987}]{kai87}
 Kaiser, N. 1987, MNRAS, 227, 1 

\bibitem[\protect\citeauthoryear{Kayo, Taruya \& Suto}{2001}]{kay01}
Kayo, I., Taruya, A., \& Suto, Y. 2001, \apj, 561, 22

\bibitem[\protect\citeauthoryear{Kim \& Strauss}{1998}]{kim98}
Kim, R. S., \& Strauss, M. A.  1998 \apj, 493, 39	

\bibitem[\protect\citeauthoryear{Kofman et al.}{1994}]{kof94} Kofman,
L., Bertshinger, E., Gelb, J. M., Nusser, A., \& Dekel, A. 1994, \apj,
420, 44

\bibitem[\protect\citeauthoryear{Lahav et al.}{1994}]{lah94} Lahav,
O., Fisher, K. B., Hoffman, Y., Scharf, C. A., \& Zaroubi, S. 1994, \apj,
423, L93

\bibitem[\protect\citeauthoryear{Lahav et al.}{2002}]{lah02}
 Lahav, O., et al. 2002,  MNRAS, 333, 961  

\bibitem[\protect\citeauthoryear{Le F\`evre et al.}{1996}]{lef96} Le
F\`evre, O., et al. 1996, \apj, 461, 534

\bibitem[\protect\citeauthoryear{Le F\`evre et al.}{2004}]{lef04} Le
F\`evre, O., et al. (the VVDS team) 2004, \aap, 428, 1043 

\bibitem[\protect\citeauthoryear{Le F\`evre et al.}{2005}]{lef05} Le
F\`evre, O., et al. (the VVDS team) 2005, \aap, 439, 845

\bibitem[\protect\citeauthoryear{Lilly et al.}{2007}]{lilf07} 
Lilly, S., et al. (the zCOSMOS team) 2007, ApJS, 172, 70

\bibitem[\protect\citeauthoryear{Linder}{2005}]{lin05} 
Linder, E. V.  2005, PhRvD, 72, 3529

\bibitem[\protect\citeauthoryear{Linder}{2007}]{lin07} 
Linder, E. V., 2007, arXiv:0709.1113

\bibitem[\protect\citeauthoryear{Marinoni et al.}{1998}]{mar98}
Marinoni, C., Monaco, P., Giuricin, G.\& Costantini, B. 1998, \apj, 505, 484

\bibitem[\protect\citeauthoryear{Marinoni et al.}{2002}]{mardav02}
Marinoni, C., Davis, M., Newmann, J. A.,  \& Coil, A. L. 2002, \apj, 580, 122

\bibitem[\protect\citeauthoryear{Marinoni \& Hudson}{2002}]{mahu02}
Marinoni, C.,  \& Hudson, M.  2002, ApJ, 569, 101 

\bibitem[\protect\citeauthoryear{Marinoni et al.}{2005}]{mar05}
Marinoni, C. et al. (the VVDS team)  2005, A\&A, 442, 801 (Paper I)

\bibitem[\protect\citeauthoryear{Massey et al.}{2007}]{mas07}
Massey, R., et al. 2007, Nature, 445, 286

\bibitem[\protect\citeauthoryear{Mc Cracken et al.}{2003}]{mcc03} 
McCracken, H., et al. 2003, \aap, 410, 17

\bibitem[\protect\citeauthoryear{Oke \& Gunn}{1983}]{oke83} 
Oke, J. B., \& Gunn, J. E.  1983, \apj, 266, 713

\bibitem[\protect\citeauthoryear{Ostriker et al.}{2003}]{ost03}
Ostriker, J. P., Nagamine, K., Cen, R., \& Fukugita, M.  2003, \apj, 597, 1

\bibitem[\protect\citeauthoryear{Peacock et al.}{2001}]{pea01}
Peacock, J. A. et al. 2001, Nature, 410, 169 

\bibitem[\protect\citeauthoryear{Peebles}{1980}]{pee80} 
Peebles, P. J. E.  1980, The Large Scale Structure of the Universe 
(Princeton: Princeton Univ Press)

\bibitem[\protect\citeauthoryear{Pollo et al.}{2005}]{pol05}
Pollo, A. et al. (the VVDS team) 2005, \aap, 439, 887

\bibitem[\protect\citeauthoryear{Pollo et al.}{2005}]{pol06}
Pollo, A. et al. (the VVDS team) 2006, \aap, 451, 409

\bibitem[\protect\citeauthoryear{Press et al.}{1992}]{pre92} Press, W. H., Teukolsky, S. A., Vetterling, W. T., \& Flannery, B. P.  1992, Numerical Recipes, (Cambridge: University Press)

\bibitem[\protect\citeauthoryear{Scaramella et al.}{1989}]{scar89}
Scaramella, R. et al. 1989, Nature, 338, 562 

\bibitem[\protect\citeauthoryear{Scherrer \& Gaztanaga}{2001}]{sg01}
Scherrer. R. J., \& Gaztanaga, E. 2001, MNRAS,  328, 257

\bibitem[\protect\citeauthoryear{Scoville et al.}{2007}]{sco07}
Scoville, N. 2007, \apjs, 172, 150

\bibitem[\protect\citeauthoryear{Sigad, Branchini \&
Dekel}{2000}]{sbd00} Sigad, Y., Branchini, E., \& Dekel, A.  2000, \apj,
540, 62

\bibitem[\protect\citeauthoryear{Somerville et al.}{2001}]{som01}
Somerville, R. S., et al. 2001, MNRAS, 320, 289  

\bibitem[\protect\citeauthoryear{Spergel et al.}{2007}]{sper07}
Spergel, D. N., et al., 2007, ApJS, 170, 377

\bibitem[\protect\citeauthoryear{Springel et al.}{2005}]{spri05}
Springel, W. et al. 2005, Nature, 435, 629 

\bibitem[\protect\citeauthoryear{Strauss et al.}{1992a}]{s92a}
Strauss, M. A., et al. 1992, ApJS, 83, 29S

\bibitem[\protect\citeauthoryear{Strauss et al.}{1992b}]{s92b}
Strauss, M. A., Davis, M.,  Yahil, A.,  \& Huchra, J. P. 1992, \apj, 385, 421

\bibitem[\protect\citeauthoryear{Strauss \& Willick}{1995}]{sw95}
Strauss, M. A., \& Willick, J. A.  1995, PhR, 261, 271

\bibitem[\protect\citeauthoryear{Taylor \& Watts}{2000}]{tay00}
Taylor, A. N., \& Watts, P. I. R.  2000, \mnras, 314, 92

\bibitem[\protect\citeauthoryear{Tegmark et al.}{2006}]{teg06}
Tegmark, M., et al. 2006, Phys. Rev. D, 74, 123507

\bibitem[\protect\citeauthoryear{Watts \& Taylor}{2000}]{wat01}
Watts, P. I. R., Taylor, A. N. 2001, \mnras, 320, 139

\bibitem[\protect\citeauthoryear{Verde et al.}{2002}]{ver02}
Verde, L. et al. 2002, MNRAS,  335, 432 

\bibitem[\protect\citeauthoryear{Wang \& Steinhardt}{1998}]{ws98}
Wang, L., \& Steinhardt, P. J. 1998, \apj, 508, 483 

\bibitem[\protect\citeauthoryear{Wild et al.}{2005}]{wild} 
Wild, V., et al. 2005, \mnras,  356, 247


\end{thebibliography}
\end{document}